\documentstyle[epsfig,pre,aps,amsfonts,amsmath]{revtex}
\draft

\makeatletter
\def\endtable{%
\global\tableonfalse\global\outertabfalse
{\let\protect\relax\small\vskip2pt\@tablenotes\par}\xdef\@tablenotes{}%
\egroup
\vskip1pc \penalty-200
}%
\makeatother

\begin{document}
\title{Novel Monte Carlo method to calculate the central charge and
critical exponents}
\author{Paul J. M. Bastiaansen\cite{paul} and Hubert J. F. Knops}
\address{Institute for Theoretical Physics,
University of Nijmegen, PO Box 9010, 6500 GL Nijmegen, The Netherlands}
\maketitle

\begin{abstract}
A typical problem with Monte Carlo simulations in statistical physics
is that they do not allow for a direct calculation of the free energy.
For systems at criticality, this means that one cannot calculate the
central charge in a Monte Carlo simulation. We present a novel finite
size scaling technique for two-dimensional systems on a geometry of
$L\times M$, and focus on the scaling behavior in $M/L$. We show that
the finite size scaling behavior of the stress tensor, the operator
that governs the anisotropy of the system, allows for a determination
of the central charge and critical exponents. The expectation value of
the stress tensor can be calculated using Monte Carlo simulations.
Unexpectedly, it turns out that the stress tensor is remarkably
insensitive for critical slowing down, rendering it an easy quantity to
simulate. We test the method for the Ising model (with central charge
$c=\frac12$), the Ashkin-Teller model ($c=1$), and the F-model (also
$c=1$).
\end{abstract}


%

\section{Introduction}

There exist basically two methods to obtain numerical information on
two-dimensional critical systems. In the transfer matrix method one
calculates the largest eigenvalue of the transfer matrix and thus one
finds the free energy of the system on a $L\times\infty$ cylinder. From
the theory of conformal invariance\cite{Cardy87} one knows that this
free energy is related to the central charge $c$ as
\begin{equation}
   f = f_\infty - \frac{ \pi c}{6L^2}.
\end{equation}
By introducing appropriate seams on the cylinder, that alter the cyclic
boundary conditions, one also obtains some of the leading critical
dimensions $x$. Here one uses the result from conformal invariance that
the central charge $\tilde{c}$ from a system with a seam is given by
\begin{equation}
   \tilde{c} = c - 12 x.
\end{equation}
After the first paper\cite{Blote86} that used this technique this
method has become very popular.

The advantage of the method is its high numerical accuracy. A distinct
disadvantage is that the method is limited to rather small values of
$L$, since the required storage capacity and computer time increase
exponentially with $L$. This also implies that the method is limited to
discrete spin systems. Both limitations are lifted but exchanged for a
loss in numerical accuracy in the Monte Carlo transfer matrix
method\cite{Nightingale88,Thijssen90}. A
promising\cite{Nishino95,Carlon97} method to study the transfer matrix
for discrete spin systems for large $L$ is density matrix
renormalization\cite{White93}.

The second method that has extensively been used to obtain numerical
information on two-dimensional critical systems is the
standard\cite{Binder84} Monte Carlo (MC) method. It can be used for
fairly large ($L\times L$) system sizes, both for discrete and
continuous spin systems. Critical exponents can be extracted from the
finite size scaling behavior of the fluctuations in critical quantities
like energy and order parameter.

However, since the free energy cannot directly be measured in a MC
simulation, there exists at the moment no MC method to evaluate the
central charge. This is unfortunate since this quantity plays such an
important role in the determination of the universality class of
two-dimensional systems.

In this paper we want to complete this palette of existing numerical
methods by presenting a direct MC method to evaluate the central
charge. We do this by constructing an operator on the lattice that in
the scaling limit represents the stress tensor $T$. The stress tensor
is an operator that is connected with the anisotropy of the system;
when one allows for anisotropy in critical models, critical points in
the phase diagram become critical lines. The whole of such a critical
line falls into the same universality class, and movements along the
line are governed by a marginal operator, having its critical dimension
$x=2$. This anisotropy operator is the stress tensor $T$, and can be
defined for any critical model. It is, in the language of conformal
invariance, the second descendant of the identity operator. The
expectation value of $T$ on a $L\times M$ torus is known from conformal
theory and contains in particular the central charge $c$. By comparing
our MC results, as a function of $M/L$, with this formula we obtain the
central charge and the leading critical dimensions.

An additional advantage of the method turns out to be that the
autocorrelation function of $T$ in the MC sequence decays much faster
than that of a critical quantity like the energy. This can only partly
be attributed to the fact that $T$ is not a relevant but only a
marginal operator. This means that there is no need to invoke more
sophisticated MC methods like that of Swendsen-Wang\cite{Swendsen87} to
avoid critical slowing down.

\section{Conformal invariance of critical field theories}
\label{cft}

Besides being invariant against a rescaling of the length parameters,
critical models are believed to be conformaly invariant as well: their
large scale behavior is invariant against transformations that
correspond locally to a rotation and a rescaling. Such transformations
are called conformal transformations. From this symmetry, present at
criticality, follows the structure of the Hilbert space for a large
part, at least in the case of two-dimensional models. We will summarize
some results that we need in the sequel; more details can be found in
the review by Cardy\onlinecite{Cardy87}.

\begin{figure}
   \begin{center}
   \epsfig{file=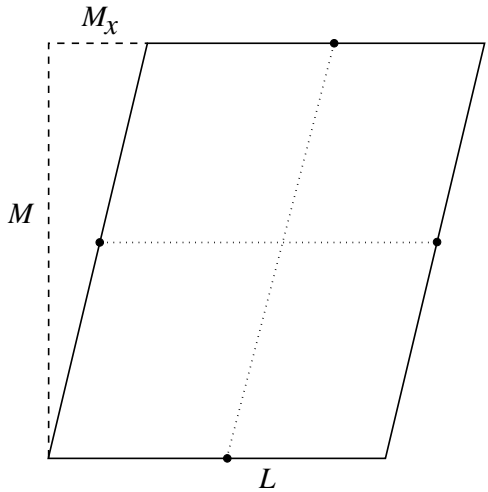,width=7cm}
   \end{center}
   \caption{The torus geometry on which the conformal field theory is
   defined. Dimensions of the torus are $L\times M$, the boundary
   conditions are such that the indicated points are identified: they
   are cyclic in the horizontal direction and cyclic with a shift over
   $M_x$ in the vertical direction.}
   \label{torus}
\end{figure}
In this section, we will be concerned with a system defined on a `skew'
torus; its dimensions are $L\times M$, and boundary conditions are
cyclic in the horizontal direction and cyclic after a shift over $M_x$
in the vertical direction, as in Fig.~\ref{torus}. Denoting the
transfer matrix of the system with $\exp(-H)$, where $H$ is the
Hamilton operator, its partition function on such a geometry is
\begin{equation}
   Z = \sum_j \big\langle j \big| e^{-MH}
      e^{iM_xP} \big| j \big\rangle,
\end{equation}
where the summation is over all configurations $|j\rangle$ in a row.
The states $|j\rangle$ make up the Hilbert space on which the transfer
matrix (or operator) $\exp(-H)$ acts. $P$ is the momentum operator, the
generator of translations in the horizontal direction. The Hamilton
operator $H$ of the model is the generator of translations in the
vertical direction and commutes with $P$. When the transfer matrix is
suitably defined, there exists an orthonormal basis of the Hilbert
space, consisting of eigenstates of the Hamiltonian and of the momentum
operator. From the theory of conformal invariance follows that these
eigenstates with their eigenvalues are closely related to the critical
dimensions of the model.

There turns out to be a set of fundamental operators present in the
theory, that are indicated as $L_n$ and $\bar{L}_n$ for $n\in\Bbb{Z}$.
They satisfy the celebrated {\sl Virasoro algebra}. The Hamiltonian $H$
and the momentum operator $P$ can be expressed in terms of $L_0$ and
$\bar{L}_0$ as follows,
\begin{equation}
\begin{split}
   H &= E_0L + \frac{2\pi}L \big( L_0 + \bar{L}_0 \big) -
      \frac{\pi c}{6L}, \\
   P &=        \frac{2\pi}L \big( L_0 - \bar{L}_0 \big).
\end{split}
\end{equation}
Here $c$ is the central charge of the model, and $E_0$ is the bulk
groundstate energy of the Hamiltonian, which is usually subtracted from
it. The eigenstates of the Hamiltonian and of the momentum operator are
labeled as $|\Delta+m,\bar\Delta+\bar{m}\rangle$, with the relations
\begin{equation}
\begin{split}
        L _0 |\Delta+m,\bar\Delta+\bar{m}\rangle &=
      (     \Delta +     m ) |\Delta+m,\bar\Delta+\bar{m}\rangle, \\
   \bar{L}_0 |\Delta+m,\bar\Delta+\bar{m}\rangle &=
      (\bar\Delta+\bar{m}) |\Delta+m,\bar\Delta+\bar{m}\rangle.
   \label{eigenvalue}
\end{split}
\end{equation}
Hence
\begin{alignat}{2}
   (H-E_0) |\Delta+m,\bar\Delta+\bar{m}\rangle &=
      \frac{2\pi}L
      \left(
         \Delta+\bar\Delta+m+\bar{m} - \frac{c}{12}
      \right) & |\Delta+m,\bar\Delta+\bar{m}\rangle,
      \label{hamiltonoperator}\\
   P |\Delta+m,\bar\Delta+\bar{m}\rangle &=
      \frac{2\pi}L
      \left(
      \Delta-\bar\Delta+m-\bar{m}
      \right) & |\Delta+m,\bar\Delta+\bar{m}\rangle.
      \label{momentumoperator}
\end{alignat}
The states $|\Delta,\bar\Delta\rangle$ with $m=\bar{m}=0$ are called
{\sl primary states} and the states with $m$ and/or $\bar{m}$ unequal
to zero are their {\sl conformal followers}. The values of $\Delta$ and
$\bar\Delta$ are related to the critical dimensions $x$ and spin
indices $l$ of the operators of the theory according to
\begin{equation}
   \begin{aligned}
      x &= \Delta + \bar\Delta + m + \bar{m} , \\
      l &= \Delta - \bar\Delta + m - \bar{m} .
   \end{aligned}
\end{equation}
The appearing values of $\Delta$ and $\bar\Delta$ from the primary
states, together with their multiplicity (the level of their
degeneracy) as well as the multiplicities of their conformal followers,
determine the full structure of the Hilbert space. The values of the
critical dimensions and their multiplicities are universal. This
implies that the partition function (divided by its bulk value that
results from $E_0$) considered as a function of $M/L$, is universal in
the scaling limit of $L$ and $M$ large.

From Eq.~(\ref{partfunction}) and Eq.~(\ref{hamiltonoperator}) follows
that the diagonal element of $\exp(-MH)\exp(iM_xP)$ for the state
$|\Delta+m,\bar\Delta+\bar{m}\rangle$ is
\begin{equation}
   Q^{-c/24} \bar{Q}^{-c/24} Q^{\Delta+m}
   \bar{Q}^{\bar\Delta+\bar{m}},
\end{equation}
with
\begin{equation}
   Q=\exp\left(
      -\frac{2\pi M}{L}+\frac{2\pi iM_x}{L}
   \right)
\end{equation}
and $\bar{Q}$ the complex conjugate of $Q$. Summing over all diagonal
elements yields the partition function of the model. Let us label the
critical dimensions with $j$, then
\begin{equation}
   Z/Z_{\text{bulk}} = Q^{-c/24} \bar{Q}^{-c/24}
   \sum_j N_j Q^{\Delta_j+m_j}
   \bar{Q}^{\bar\Delta_j+\bar{m}_j}.
   \label{partfunction}
\end{equation}
This expression is called the {\sl universal expression for the
partition function}, and contains the central charge $c$, the values of
the critical dimensions $\Delta_j+m_j$ and $\bar\Delta_j+\bar{m}_j$ as
well as their multiplicities $N_j$. In the limit
$M/L\rightarrow\infty$, it yields the well known finite size scaling
relation for the central charge, used in transfer matrix calculations,
\begin{equation}
   f(L) = f(\infty) - \frac{\pi c}{6L^2}
\end{equation}
where
\begin{equation}
   f(L) = -\lim_{M/L\rightarrow\infty} \frac1{ML}\ln(Z).
\end{equation}
The free energy, however, is not directly accessible in MC simulations.
Consequently there exist no MC results that yield the central charge of
critical models. The stress tensor, however, is an operator that is
closely related to the Hamilton operator and it this operator (or
rather its expression in terms of spin variables) that actually {\sl
is} accessible in MC simulations, in contrast to the free energy. We
will show this in the sequel.

Conformal invariance in critical field theories states that the action
(or in statistical mechanics terms: the classical interaction) is
invariant against conformal transformations. The change in the action
for non-conformal transformations is determined by the stress tensor
$T({\bf r})$,
\begin{equation}
   T({\bf r}) = \begin{pmatrix}
          T_{xx}({\bf r}) & T_{xy}({\bf r}) \\
          T_{yx}({\bf r}) & T_{yy}({\bf r})
       \end{pmatrix},
\end{equation}
where $T({\bf r})$ is a symmetric, traceless tensor. Hence $T_{xx}({\bf
r})=-T_{yy}({\bf r})$ and $T_{xy}({\bf r})=T_{yx}({\bf r})$. Usually,
one defines the independent components of $T$ as
\begin{equation}
\begin{split}
        T (u,v) &= \frac12 \big[ T_{xx}(u,v) - i T_{xy}(u,v) \big],\\
   \bar{T}(u,v) &= \frac12 \big[ T_{xx}(u,v) + i T_{xy}(u,v) \big].
\end{split}
\end{equation}
Here $u$ and $v$ are the position coordinates on the torus. The
dimension $(\Delta,\bar\Delta)$ of $T$ and $\bar T$ are $(2,0)$ and
$(0,2)$ respectively. Hence their critical dimension $x=2$ and their
spin indices are $l=\pm2$. So the stress tensor is a marginal operator.
Its components $T$ and $\bar{T}$ can be expressed in terms of the
fundamental Virasoro operators $L_n$ and $\bar{L}_n$ as follows:
\begin{equation}
\begin{split}
   T(u,v) &= \left(\frac{2\pi}L\right)^2
      \left(
      \frac{c}{24} - \sum_{n=-\infty}^{\infty}
      e^{-\frac{2\pi i}L un} e^{\frac{2\pi }L vn} L_n \right), \\
   \bar{T}(u,v) &= \left(\frac{2\pi}L\right)^2
      \left(
      \frac{c}{24} - \sum_{n=-\infty}^{\infty}
      e^{+\frac{2\pi i}L un} e^{\frac{2\pi }L vn} \bar{L}_n \right).
   \label{stressdef}
\end{split}
\end{equation}
This expression is valid on the torus geometry; $u$ ($v$) is the
horizontal (vertical) position on the torus. The Virasoro operators
$L_n$ and $\bar{L}_n$ with $n\neq 0$ play the role of raising and
lowering operators for the states
$|\Delta+m,\bar\Delta+\bar{m}\rangle$. Because these states are
orthonormal, only $L_0$ and $\bar{L}_0$ have nonvanishing contributions
in the expression for the expectation value of $T$.
From the expression~(\ref{stressdef}) and the eigenvalue
equations~(\ref{eigenvalue}), the expression for the expectation value
of the stress tensor can easily be calculated. The expectation value is
\begin{equation}
   \langle T\rangle = \frac1Z \sum_j
   \big\langle\Delta_j+m_j,\bar\Delta_j+\bar{m}_j | \, T \,
   |\Delta_j+m_j,\bar\Delta_j+\bar{m}_j\big\rangle.
\end{equation}
Substituting Eq.~(\ref{stressdef}) and the expression for the
partition function~(\ref{partfunction}) yields
\begin{equation}
\begin{split}
   \langle T\rangle &=
   \left(\frac{2\pi}{L}\right)^2
   \left( \frac{c}{24} -
     \dfrac{
     \sum_j N_j \left( \Delta_j+m_j\right)
            Q^{-\Delta_j-m_j} \bar{Q}^{-\bar\Delta_j-\bar{m}_j}
   }{
     \sum_j N_j Q^{-\Delta_j-m_j} \bar{Q}^{-\bar\Delta_j-\bar{m}_j}
   }
   \right), \\[1.3ex]
   \langle \bar{T}\rangle &=
   \left(\frac{2\pi}{L}\right)^2
   \left( \frac{c}{24} -
     \dfrac{
     \sum_j N_j \left( \bar\Delta_j+\bar{m}_j\right)
            Q^{-\Delta_j-m_j} \bar{Q}^{-\bar\Delta_j-\bar{m}_j}
   }{
     \sum_j N_j Q^{-\Delta_j-m_j} \bar{Q}^{-\bar\Delta_j-\bar{m}_j}
   }
   \right).
\end{split}
\end{equation}
We will be mostly concerned with the diagonal elements $T_{xx}=-T_{yy}$
of the stress tensor. The expression for $T_{xx}$ is
\begin{equation}
   \langle T_{xx}\rangle =
   \left(\frac{2\pi}{L}\right)^2
   \left( \frac{c}{12} -
      \dfrac{
        \sum_j N_j\left(\Delta_j+m_j+\bar\Delta_j+\bar{m}_j\right)
        Q^{-\Delta_j-m_j} \bar{Q}^{-\bar\Delta_j-\bar{m}_j}
      }{
        \sum_j N_j Q^{-\Delta_j-m_j} \bar{Q}^{-\bar\Delta_j-\bar{m}_j}
      }
   \right).
   \label{stressxx}
\end{equation}
This expectation value can be written as the derivative of the free
energy with respect to the aspect ratio $M/L$, as
\begin{equation}
   \langle T_{xx}\rangle = -2\pi \frac{M}L
      \frac{\partial f}{\partial (M/L)}.
   \label{partderivative}
\end{equation}
Upon taking the derivative, the volume $ML$ of the system is kept
constant. Like magnetization and magnetic field, the `field' $M/L$ is
the external field conjugate to the operator that is the stress
tensor. Therefore, $\langle T_{xx}\rangle$ couples to the anisotropy
of the system.

\section{A lattice representation of the stress tensor}
\label{discrete}

In conformal field theory, the stress tensor is an operator that is
quite abstract. It is defined only after the lattice model has gotten
its continuum limit. For lattice models, however, the stress tensor can
easily be defined as well. This lattice representation of the stress
tensor must thus have the scaling behavior predicted by
expression~(\ref{stressxx}). Below we will illustrate the construction
of the lattice representation of the stress tensor for the Ising model,
but first we give a more general way to proceed.

\subsection{Constructing the stress tensor}

Construct, for a lattice model, an operator $t({\bf r})$ as an
expression in the local, fluctuating field(s), such that: (i) $t({\bf
r})$ transforms as a second rank tensor (in particular, $t({\bf r})$
picks up a minus sign under a rotation over 90$^\circ$); and (ii)
$t({\bf r})$ has the same symmetry as the interaction energy of the
model under study. In general, this means that $t({\bf r})$ is
invariant under global spin flips or spin rotations.

If one now expresses $t({\bf r})$ in terms of scaling operators it is
clear that the operators that occur should all be tensors that change
sign under rotations over $\pi/2$, i.e., they have
$l=\Delta-\bar\Delta=\pm2, \pm6,\ldots$. Since $\Delta$ and
$\bar\Delta$ are always non-negative, it follows that the scaling
dimension $x$ of the appearing scaling operators all have $x\geq2$. The
marginal case, having $x=2$ and $l=\pm2$, is in fact the stress tensor,
all other operators in the expansion are irrelevant. To be more
precise: in this general case, the operator $t({\bf r})$ couples to
both independent components $T_{xx}({\bf r})$ and $T_{xy}({\bf r})$ of
the stress tensor. As will become clear below, however, $t({\bf r})$
can easily be defined such that it couples to $T_{xx}({\bf r})$ only.
In that case, one has
\begin{equation}
   t({\bf r}) = \alpha\; T_{xx}({\bf r}) + \cdots,
   \label{stressexpansion}
\end{equation}
where the dots represent irrelevant operators. The requirement (ii)
guarantees that $t({\bf r})$ and $T_{xx}({\bf r})$ share the same
interaction symmetry, so that the coefficient $\alpha$ does not vanish
by symmetry. Constructing operators $t({\bf r})$ can, as we shall see,
be done in several ways, but all choices yield an
expansion~(\ref{stressexpansion}), albeit with different values of
$\alpha$.

Having constructed the operator $t({\bf r})$ one can evaluate its
average in a MC simulation on a geometry of $L\times M$, for several
values of $M/L$ and $L$ large. The result should follow the universal
expression for $T_{xx}({\bf r})$ as
\begin{equation}
   \langle t({\bf r}) \rangle =
   \alpha\langle T_{xx}({\bf r}) \rangle + {\cal O}(\frac1{L^\omega}),
\end{equation}
where the expression for $\langle T_{xx}({\bf r}) \rangle$ given in
Eq.~(\ref{stressxx}) is proportional to $1/L^2$ and dominates the
second term that has $\omega>2$. Hence we can fit the $M/L$ dependence
of the left hand side against Eq.~(\ref{stressxx}), obtaining, in
particular, the central charge $c$.

\subsection{The stress tensor for the Ising model}

We will illustrate the construction of the discrete stress tensor
$t({\bf r})$ in the case of the Ising model. The starting point is the
close connection between stress tensor and anisotropy. Let us therefore
start with the anisotropic action $\cal A$ of the ordinary, square
lattice Ising model,
\begin{equation}
   {\cal A} = - \sum_{ij}
	  \Big( J_x\; S_{i,j} S_{i+1,j} +
            J_y\; S_{i,j} S_{i,j+1} \Big),
   \label{isinghamiltonian}
\end{equation}
where the couplings $(J_x,J_y)$ allow for anisotropy. The isotropic
critical point is $J_x = J_y = J_c = \frac12\ln(1+\sqrt2)$, but this
point becomes a critical line when unequal values of $J_x$ and $J_y$
are allowed for.

The central notion here is that in the scaling limit, anisotropy
amounts to a rescaling of the length parameters $x$ and $y$ with a
different scaling factor. Hence, in the scaling limit, the anisotropic
model with $(J_x,J_y)$ behaves as the isotropic model with rescaled
length parameters $x$ and $y$,
\begin{equation}
   \begin{aligned}
      x' &= e^{-\lambda}x , \\
      y' &= e^{+\lambda}y.
   \end{aligned}
   \label{lengthrescaling}
\end{equation}
The value of $\lambda$ in this equation determines the values of $J_x$
and $J_y$. In this way, Eq.~(\ref{lengthrescaling}) fixes the
parameterization $[J_x(\lambda),J_y(\lambda)]$ of the critical line
with $\lambda$. The isotropic point has $\lambda=0$ with
$J_x(0)=J_y(0)=J_c$, and the parameterization obeys
$J_x(\lambda)=J_y(-\lambda)$.

On a finite geometry $L\times M$, this anisotropic rescaling means that
the volume $ML$ of the system remains untouched, but that the aspect
parameter $M/L$ scales according to
\begin{equation}
   \frac{M'}{L'} = e^{2\lambda} \frac{M}L .
\end{equation}
In the scaling limit, therefore, the partition function with the
anisotropic action of equation~(\ref{isinghamiltonian}), which we call
$Z(\lambda,M/L)$ depending on $\lambda$, equals that of the isotropic
Hamiltonian with a rescaled aspect ratio $M'/L'$,
\begin{equation}
   Z\left(\lambda,\frac{M}L\right) =
      Z\left(\lambda=0,e^{2\lambda} \frac{M}L\right).
   \label{scaledpartfunc}
\end{equation}

A general movement in the phase diagram is performed by a scaling
operator. A renormalization transformation is isotropic, which implies
that there can be no renormalization flow along the critical line
$(J_x,J_y)$. This implies that the scaling operator that governs the
movement along this line must be invariant against a renormalization
transformation, i.e., it is a marginal operator having its critical
dimension $x=2$.

In the case of the Ising model, the action~(\ref{isinghamiltonian})
immediately shows which operator this must be. Write the action as a
symmetric part plus a part that determines the anisotropy,
\begin{equation}
   {\cal A} = {\cal A}_c -
	  \sum_{ij} (J_x(\lambda)-J_c)\, S_{i,j}S_{i+1,j} +
                (J_y(\lambda)-J_c)\, S_{i,j}S_{i,j+1},
\end{equation}
where ${\cal A}_c$ is the action at the isotropic critical point
$J_x=J_y=J_c$. Expanding up to first order in $\lambda$, this
expression can be written as
\begin{equation}
   {\cal A} = {\cal A}_c + \lambda \sum_{ij} t_{xx}(i,j),
\end{equation}
where $t_{xx}(i,j)$ is the lattice representation of the stress tensor,
\begin{equation}
   t_{xx}(i,j) = - J_x'(0)
      \Big( S_{i,j}S_{i+1,j} - S_{i,j}S_{i,j+1} \Big).
   \label{latticestress}
\end{equation}
Here we used the symmetry property $J_x(\lambda)=J_y(-\lambda)$. The
operator $t_{xx}(i,j)$ governs the anisotropy of the system. This
lattice representation of the stress tensor~(\ref{latticestress}) for
the Ising model was already known for a long time\cite{Kadanoff71}; the
value of $J_x'(0)=\frac12\sqrt{2}$.

In fact, the operator in Eq.~(\ref{latticestress}) is written as
$t_{xx}$ because it is one of the two components of the full stress
tensor $t_{\alpha\beta}(i,j)$. This $t_{\alpha\beta}(i,j)$ has the same
properties as the field theoretical stress tensor: it is a second-rank,
symmetric traceless tensor. The other component, $t_{xy}(i,j)$ can be
written as
\begin{equation}
   t_{xy}(i,j) = -J_{xy}
      \Big( S_{i,j}S_{i+1,j+1} - S_{i,j}S_{i+1,j-1} \Big),
\end{equation}
with a certain prefactor $J_{xy}$, which will be different from
$J_x'(0)$, because it couples next-nearest neighbor spins instead of
nearest neighbors. The off-diagonal elements of the discrete stress
tensor couple to the anisotropy in the diagonal directions.

It is this operator $t_{\alpha\beta}({\bf r})$ that appears in the
previous subsection. It is constructed such that it behaves as a
second-rank symmetric tensor with the same symmetry under global spin
flips as the interaction energy itself. Of course, this version of
$t_{\alpha\beta}(i,j)$ is not the only possible one; it can also be
defined with further neighbor interactions.

The precise connection between the discrete variant $t({\bf r})$ of the
stress tensor and its field theoretical counterpart is obtained by
taking the derivative of Eq.~(\ref{scaledpartfunc}) with respect to
$\lambda$ at $\lambda=0$. Using Eq.~(\ref{partderivative}), this yields
\begin{equation}
    J_x'(0) \sum_{ij} \Big\langle S_{i,j}S_{i+1,j} -
       S_{i,j}S_{i,j+1}\Big\rangle
       = \frac{ML}{\pi} \; \big\langle T_{xx}(u,v) \big\rangle,
    \label{combine}
\end{equation}
where $\langle T_{xx}(u,v)\rangle$ is the expression~(\ref{stressxx})
for the expectation value of the diagonal component of the stress
tensor. Note that this expression is a universal function of central
charge, critical dimensions, and their multiplicities. The value of
$J_x'(0)$, however, is in general unknown, such that we will have to
include it as a fit parameter.

Expression~(\ref{combine}) combined with Eq.~(\ref{stressxx}) yields
the relation that is central to this work: it expresses the expectation
value of the lattice representation of the stress tensor in the
universal quantities that we want to know. As we will use a rectangular
geometry, without the shift in boundary conditions, we put $M_x=0$ and
obtain
\begin{equation}
   \big\langle S_{i,j}S_{i+1,j}-S_{i,j}S_{i,j+1}\big\rangle  =
   \alpha \left(\frac{2\pi}{L}\right)^2
   \left( \frac{c}{12} -
      \dfrac{
        \sum_j N_j \; x_j \exp\left(-2\pi\frac{M}L x_j\right)
      }{
        \sum_j N_j \exp\left(-2\pi\frac{M}L x_j\right)
     }
   \right),
   \label{fitexpression}
\end{equation}
\begin{figure}
   \begin{center}
   \epsfig{file=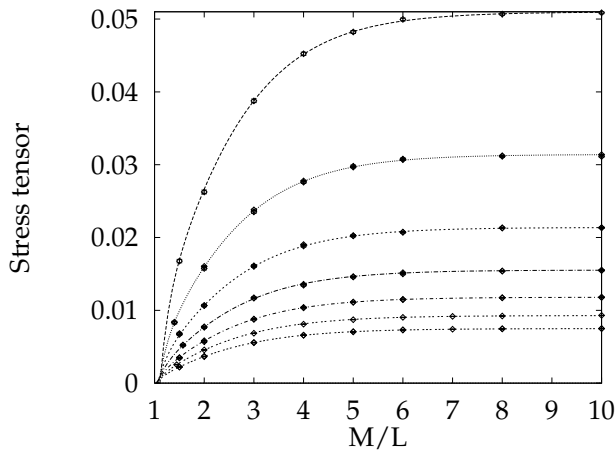,width=8.6cm}
   \end{center}
   \caption{The expectation values of stress tensor (1) for the Ising
   model, defined in Eq.~(\ref{isingstress}), as a function of the
   aspect parameter $M/L$. From high to low, the plots show the
   expectation values from system dimensions $L$ running from 4 to 10.
   The lines are the result of the fit against
   Eq.~(\ref{fitexpression}) together with a correction to scaling term
   (\ref{corrections}). Note that, for each value of $L$, the stress
   tensor at $M/L=1$ is zero by symmetry.}
   \label{isingresult}
\end{figure}

where $1/\alpha=\pi J_x'(0)$. See Fig.~\ref{isingresult} for an example
of the functional dependence. The prefactor $\alpha$ is the same
$\alpha$ that appears in Eq.~(\ref{stressexpansion}). The physical
interpretation of it is given by the relation $1/\alpha=\pi J_x'(0)$;
it determines the `amount of anisotropy' that the system obtains, once
the stress tensor is switched on. Note that $\alpha$ is non-universal;
it depends on the precise definition of the model, as well as on the
definition of the stress tensor. The other quantities present in
Eq.~(\ref{fitexpression}), however, are the central charge, the
critical dimensions, and their multiplicities, and those are all
universal. There is an infinite number of critical dimensions, but only
a limited number of these is `small', say, less than two. For large
enough values of the aspect ratio $M/L$ only a limited number of
critical dimensions have nonvanishing contributions to
Eq.~(\ref{fitexpression}) such that a fit of this expression against MC
data must be feasible. Typically, we will take $M/L\gtrsim1$.

\section{Fitting the Monte Carlo results}

Fitting the MC results against the universal expression for the stress
tensor~(\ref{fitexpression}) requires a decent fit program, as the
number of fit parameters is quite large, and requires some theoretical
reflection on the model as well. We will deal with the use of the
universal expression~(\ref{fitexpression}) and the corrections to
scaling in different subsections.

\subsection{The universal expression for the stress tensor}

The number of critical dimensions $x_j$ that appears in the universal
expression~(\ref{fitexpression}) for the stress tensor is infinite,
which clearly is an infeasible number of fit parameters. Most of the
dimensions, however, are large. Their contribution to
Eq.~(\ref{fitexpression}) goes as $\exp(-2\pi x M/L )$, so if we limit
the calculations to values of the aspect ratio $M/L$ that are not too
small, most of the dimensions $x_j$ have a vanishing contribution.
Performing some preliminary MC simulations suggests a reasonable lower
bound to $M/L$. Typically, we took $M/L \gtrsim 1$. The upper bound on
$M/L$ is determined by the value of $M/L$ where $\langle t_{xx}({\bf
r})\rangle$ reaches its asymptotic value. To determine a reasonable
upper bound on $M/L$, the same preliminary simulations can be used. The
asymptotic value of $\langle t_{xx}({\bf r})\rangle$ is
\begin{equation}
   \langle t_{xx}({\bf r})\rangle\big|_{M/L\rightarrow\infty} =
      \alpha \left( \frac{2\pi}L\right)^2 \frac{c}{12}.
\end{equation}
This expression also shows why the asymptotic value itself is not
sufficient for the determination of the central charge: it only gives
an estimate of $\alpha c$ instead of $c$.

Performing simulations to obtain the expectation value of the stress
tensor between these bounds on $M/L$ is in principle sufficient to
extract the desired quantities by fitting the results against
expression~(\ref{fitexpression}). Typically, we take three or four
critical dimensions into account, the identity dimension $x_0 = 0$,
present in any critical model, and two or three nontrivial ones $x_j$.
For these dimensions, the multiplicities $N_j$ must be specified in the
expression as well.

A multiplicity is always integer, and when it is larger than one, the
corresponding critical dimension $x$ is degenerate, and thus an
additional symmetry is present in the model. Some theoretical
reflection on the model often is sufficient to reveal such symmetries.
Another possibility is to perform fits with different values of the
multiplicities and to choose the set that gives the best fit.

The lowest appearing values of the dimensions $x=\Delta+\bar\Delta$
correspond to primary fields. As noted in Sec.~\ref{cft}, to each
primary field belongs a tower of conformal followers or descendants,
that have values of the critical dimensions that differ by an integer
from that of the primary field; they are $\Delta+m$ and
$\bar\Delta+\bar{m}$ with $m,\bar{m}\in\mathbb{N}$. The first
descendant of a scalar primary field $O({\bf r})$ with dimensions
$(\Delta,\Delta)$ is $\nabla O({\bf r})$, a vector field that has two
components; one having $(\Delta+1,\Delta)$ and the other having
$(\Delta,\Delta+1)$. To improve the fit, we will include this first
descendant into Eq.~(\ref{fitexpression}). This inclusion introduces no
new fit parameters; the value of its critical dimension is $x+1$ when
$x$ is the critical dimension of the primary field, and this value
appears twice. Hence the multiplicity of the first descendant is twice
that of the corresponding primary field.

Having fixed the multiplicities in expression~(\ref{fitexpression}),
this leaves us with four or five free parameters: the prefactor
$\alpha$, the central charge $c$, and two or three nontrivial
dimensions~$x_j$.

\subsection{Corrections to scaling}

The above analysis of the universal behavior of the stress tensor is
valid in the scaling limit. The discrete version of the stress tensor
$t_{xx}({\bf r)}$, however, is not a scaling field; as argued in
Sec.~\ref{discrete} it can in fact be written as the
expansion~(\ref{stressexpansion}) in scaling fields, of which only the
first term is the true stress tensor with its universal behavior. The
fit to this expression is treated in the previous subsection, but for
smaller system sizes other terms in the expansion become important. The
scaling behavior of these terms in $L$ goes as $L^{-\omega}$ with
$\omega>2$.

To obtain accurate results, we should include at least one of these
correction terms in the expression that we fit against our MC results.
That means that we have to perform calculations for different values of
the width $L$ of the system in order to be able to extract the $L^{-2}$
behavior of the true stress tensor.

In principle, we could proceed by performing simulations for a fixed
value of $M/L$ and increasing values of $L$ and extract, by
extrapolation, the part of $\langle t_{xx}({\bf r})\rangle$ that scales
as $L^{-2}$. This means that, for any value of $M/L$, we have to fit
\begin{equation}
   \langle t_{xx}({\bf r})\rangle = \frac{a}{L^2} + \frac{b}{L^\omega},
\end{equation}
and have to use the values of $a$ for each value of $M/L$ to fit
against expression~(\ref{fitexpression}). We can, however, do better.

To this end, write the first scaling operator on the dots in
expansion~(\ref{stressexpansion}) as $O({\bf r})$,
\begin{equation}
   \langle t_{xx}({\bf r})\rangle =
      \alpha\;\langle T_{xx}({\bf r})\rangle +
			  \langle O({\bf r})\rangle.
\end{equation}
Now consider the general expression for the expectation value of an
operator $O({\bf r})$ on a system with geometry $L\times M$,
\begin{equation}
   \langle O({\bf r})\rangle =
      \frac1Z \sum_j \langle j | \; e^{-MH} O({\bf r})\; | j \rangle,
\end{equation}
with $Z$ the partition function and $H$ the Hamilton operator of
Eq.~(\ref{hamiltonoperator}). Using the basis
$|\Delta+m,\bar\Delta+\bar{m}\rangle$ of the Hilbert space yields
\begin{equation}
   \langle O({\bf r}) \rangle =
     \dfrac{
        \sum_j N_j a_j(L) e^{-2\pi \frac{M}L x_j}
     }{
        \sum_j N_j e^{-2\pi \frac{M}L x_j}
     }
   \label{expectvalue}
\end{equation}
where we used Eq.~(\ref{hamiltonoperator}). Here the parameters
$a_j(L)$ are the diagonal elements of the operator $O({\bf r})$ in the
basis $|\Delta+m,\bar\Delta+\bar{m}\rangle$ of the Hilbert space,
\begin{equation}
   a_j(L) = \langle\Delta_j+m_j,\bar\Delta_j+\bar{m}_j |
      \; O({\bf r}) \; | \Delta_j+m_j,\bar\Delta_j+\bar{m}_j \rangle.
\end{equation}
As the basis functions $|\Delta+m,\bar\Delta+\bar{m}\rangle$ only
depend on $L$ and not on $M/L$, the full $M/L$ dependence of the
expectation value $\langle O({\bf r})\rangle$ is accounted for by the
exponentials in Eq.~(\ref{expectvalue}). The amplitudes $a_j(L)$ depend
on $L$ only. Taking only the leading correction into account, they can
be written as
\begin{equation}
   a_j(L) = \frac{a_j}{L^\omega},
   \label{amplitudes}
\end{equation}
with the same value of $\omega$ for each of the diagonal
elements\cite{Cardy87}. In our fit, we will only include the few most
important critical dimensions $x_j$: the identity dimensions $x_0=0$
and the first two nontrivial ones. That means that including the
expression for $\langle O({\bf r})\rangle$ as a correction to scaling
gives only four additional fit parameters: the correction exponent
$\omega$ and three amplitudes $a_0$, $a_1$, and $a_2$.

In the other case, by naively extrapolating the behavior of $\langle
t_{xx}({\bf r})\rangle$ for large $L$, we have to include the two
fit parameters $\omega$ and $b$ for each value of $M/L$. The above
approach thus drastically reduces the number of fit parameters.

Still, in the complete analysis of the MC results, the number of
fit parameters is quite large. Typically, we need four parameters from
the expression~(\ref{fitexpression}), which are the prefactor $\alpha$,
the central charge $c$, and the two most relevant dimensions $x_1$ and
$x_2$. For the corrections to scaling, we use the expression following
from Eq.~(\ref{expectvalue}) and (\ref{amplitudes}),
\begin{equation}
   \langle O({\bf r}) \rangle = \frac1{L^\omega}
     \dfrac{
        \sum_j N_j a_j \; e^{-2\pi \frac{M}L x_j }
     }{
        \sum_j N_j e^{-2\pi \frac{M}L x_j }
     }
   \label{corrections}
\end{equation}
giving four additional parameters, which are $\omega$, $a_0$, $a_1$ and
$a_2$.

In this way, we perform a combined fit of the MC results for all values
of $L$ and $M/L$, in one single fit using eight fit parameters. This
is a large number, but the functional dependence of the formula, to be
fitted for two variables simultaneously, is very restrictive.
Especially for expression~(\ref{fitexpression}), the behavior in $L$ is
restricted to $L^{-2}$, and the values of the dimensions $x_1$ and
$x_2$ appear as dimension as well as as amplitude.

The number of critical dimensions $x_j$ and the values of $M/L$ that
have to be included in the fit are a matter of trial and error.
Sometimes it turned out necessary to delete some of the lowest values
of $M/L$ from the data set. Lower values of $M/L$ clearly stabilize the
fit, but on the other hand, including these values requires more
critical dimensions from expression~(\ref{fitexpression}) to describe
the full data set. We varied the lower bound on $M/L$ and the number of
critical dimensions, until the quality of the fit became high enough.

This procedure requires a fit program that yields, apart from the
values of the fit parameters and their error bars, also a parameter
that indicates whether the fit can be trusted or not. Our case amounts
to a two-dimensional fit (in $L$ and $M/L$) using eight or ten 
fit parameters. The program we used is based on routines from {\sc
Numerical Recipes}\cite{numrec}. The parameter that indicates the
quality of the fit is called the {\sl goodness of fit} $Q$. The value
of $Q$ lies between 0 and 1, and is based on the $\chi^2$ of the fitted
data. $Q$ gives the probability that the $\chi^2$ of a certain data set
exceeds that of the actual data set. A very low value of $Q$ means that
it is highly unlikely that the used function gives the correct
theoretical description of the data. In our case this means that we
either included values of $M/L$ that are too small, or not enough
critical dimensions $x_j$.

\section{Comparison with exactly solved models}

In order to test the method, we performed MC simulations on some models
of which the scaling behavior on the torus is known exactly. We chose
the Ising model (with central charge $c=\frac12$), the Ashkin-Teller
model (with $c=1$) and the F-model (also with $c=1$). There is a line
in the phase diagram of the Ashkin-Teller model that can be mapped, by
a duality transformation and a graphical representation\cite{Baxter82},
exactly on the F-model. We chose to simulate the corresponding points
in the Ashkin-Teller model and the F-model. The results, however,
differ, which is an illustration of the importance of boundary
conditions in such simulations. The duality transformation alters the
boundary conditions, giving rise to a different behavior of both models
on a finite geometry.

In case of the Ising and Ashkin-Teller models, we performed MC
simulations using the standard Metropolis algorithm. For the F-model,
we had to use a cluster algorithm as well (to be described below). We
performed simulations on a system with geometry $L\times M$ with
varying values of $L$ as well as of $M/L$. We sampled different
versions of the stress tensor $t_{xx}({\bf r})$, in order to obtain
independent estimates of central charge and critical dimensions.

\subsection{The Ising model}

We carried out our simulations on the ordinary square lattice Ising
model, with the action given in Eq.~(\ref{isinghamiltonian}) on its
isotropic critical point given by
$J_x=J_y=J_c=\frac12\ln(1+\sqrt{2})$. The construction of the stress
tensor is described in Sec.~\ref{discrete}. Actually, taking different
versions of the discrete stress tensor $t_{xx}({\bf r})$ gives an
independent check on the accuracy of the results. All different
versions should couple to the true stress tensor $T_{xx}({\bf r})$,
albeit with different prefactors $\alpha$. We chose two versions of the
discrete stress tensor, one defined with nearest-neighbor couplings and
the other with next-next-nearest neighbor couplings:
\begin{equation}
   \begin{aligned}
   (1) \quad& \langle S_{i,j}S_{i+1,j} - S_{i,j}S_{i,j+1}\rangle, \\
   (2) \quad& \langle S_{i,j}S_{i+2,j} - S_{i,j}S_{i,j+2}\rangle.
   \end{aligned}
   \label{isingstress}
\end{equation}
Note that the stress tensor defined with next-nearest neighbor
couplings corresponds to the off-diagonal elements of the stress
tensor; its expectation value on the used geometry is zero by
symmetry. We took the system geometry $L\times M$ with $L$ varying
from 4 to 10 and $M/L$ varying from 1.5 to 10.

\begin{table}
   \caption{Monte Carlo results for the Ising model. Stress tensors (1)
   and (2) refer to the definition in Eq.~(\ref{isingstress}). Values
   of the prefactor $\alpha$, central charge $c$ and the first two
   critical dimensions $x_1$ and $x_2$ are given and compared with
   their exact values. Errors in the last digit are given between
   parentheses. $\omega$ is the power of the $1/L$ correction, and
   g.o.f.\ is the `goodness of fit'. In case of stress tensor (1), the
   prefactor $\alpha$ is known exactly\protect\cite{Kadanoff71}.}
\begin{tabular}{cd@{~}ld@{~}ll}
     &  \multicolumn{2}{c}{Stress tensor (1)} &
        \multicolumn{2}{c}{Stress tensor (2)} & Exact \\
 \tableline
   $\alpha$ &  0.450   & (2) &  1.277  & (3) & $\sqrt2/2 = 0.4501...
      $\tablenote{only for stress tensor (1)} \\
   $c$      &  0.500   & (2) &  0.498  & (1) & 1/2              \\
   $x_1$    &  0.1254  & (6) &  0.1256 & (4) & 1/8              \\
   $x_2$    &  1.0     & (4) &  1.1    & (4) & 1                \\
   $\omega$ &  4.3     & (1) &  4.29   & (8) &                  \\
   g.o.f.   &  0.83    &     &  0.97   &     &                  \\
\end{tabular}
\label{table_ising}
\end{table}

The resulting expectation values were fitted against
expression~(\ref{fitexpression}) together with a correction to scaling
term of Eq.~(\ref{corrections}). We took two nontrivial critical
dimensions $x_1$ and $x_2$ into account, both with multiplicity~1. The
data for stress tensor (1), together with the results of our fit, are
plotted in Fig.~\ref{isingresult} to get a feeling of the behavior of
the stress tensor. The numerical results of the fit are summarized in
table~\ref{table_ising}. Even for those small system sizes, accurate
results are obtained. This is the first determination of the central
charge using a MC simulation.

\subsection{The Ashkin-Teller model}

A more severe test of the method is obtained by considering a model
having dimensions lying closer to each other. The Ashkin-Teller model
is a useful candidate for testing our method. It has in its phase
diagram a critical line which can be mapped on the (exactly solved) six
vertex model\cite{Baxter82}. The universal partition sum of the
Ashkin-Teller model on the torus is exactly known\cite{Saleur87}, so it
can be compared with our MC results.

The Ashkin-Teller model has two Ising spins $S$ and $P$ with
$S,P\in\{+1,-1\}$ on each lattice site, that interact with an action
\begin{equation}
   {\cal A} = - \sum_{\langle ij\rangle}
      J \left( S_iS_j + P_iP_j \right) +
      K S_iS_jP_iP_j,
   \label{AThamiltonian}
\end{equation}
where $\langle ij\rangle$ denotes a summation over nearest neighbor
lattice sites. The critical line in the phase diagram that can be
mapped on the six vertex model is parameterized by
\begin{equation}
\begin{aligned}
   \exp(2J)    &= \frac{1-W}W, \\
   \exp(2J+2K) &= \frac{1+W}{1-W}.
\end{aligned}
\label{mapping}
\end{equation}
The weight $W$ equals the Boltzmann weight of the four vertices in the
six vertex model that carry a step. The other vertices are flat and
have Boltzmann weight~1.

The critical line of Eq.~(\ref{mapping}) is a line with central charge
$c=1$ and continuously varying exponents. By expressing the partition
function of the Ashkin-Teller model in the scaling limit in terms of
Coulomb gas partition functions, all critical exponents can be
obtained. For this derivation, the reader is referred to
Ref.~\onlinecite{Saleur87}; we will only state the results.

Part of the exponents varies continuously along the critical line.
Their value is expressed in terms of the renormalized value of the
Gaussian coupling $g$, present in the Coulomb gas partition functions.
The dimensions $x$ of the primary fields are
\begin{equation}
   x = \frac{e^2}{2g} + \frac{gm^2}2
      \quad\text{with $e,m\in\mathbb{Z}$},
   \label{cgp}
\end{equation}
and $g$ is the Gaussian coupling, given by
\begin{equation}
   g = \frac8\pi \arcsin\left( \frac1{2W} \right).
\end{equation}
The other dimensions are constant along the critical line. We chose,
rather arbitrarily, the point $W = 0.8$ on the critical line for our
simulations. At this point, the three most relevant dimensions are
\begin{equation}
\begin{aligned}
   x_1 &= 0.125     &\quad\text{(with multiplicity 2)}& ,\\
   x_2 &= 0.2908... &\quad\text{(with multiplicity 1)}& ,\\
   x_3 &= 0.8596... &\quad\text{(with multiplicity 1)}& .
\end{aligned}
\end{equation}
Typically, the multiplicity of the degenerate dimension $x_1$ (which is
constant along the critical line) can be guessed beforehand, albeit
some theoretical reflection on the model is necessary. To this end,
consider the expansion in scaling operators of $S$ and $P$,
\begin{equation}
\begin{aligned}
   S({\bf r}) &= \alpha_S \; p({\bf r}) + \cdots, \\
   P({\bf r}) &= \alpha_P \; q({\bf r}) + \cdots,
\end{aligned}
\end{equation}
where $p({\bf r})$ and $q({\bf r})$ are the leading (most relevant)
scaling operators in the expansion. The manifest symmetry $S
\leftrightarrow P$ of the action~(\ref{AThamiltonian}) implies that
\begin{equation}
   \alpha_S^2 \; \langle p({\bf r}_1) p({\bf r}_2) \rangle =
   \alpha_P^2 \; \langle q({\bf r}_1) q({\bf r}_2) \rangle.
\end{equation}
Hence it follows that $p({\bf r})$ and $q({\bf r})$ share the same
critical dimensions $x$. On the other hand, spin reversal symmetry
$S\rightarrow -S$ implies that
\begin{equation}
   \langle S({\bf r}_1) P({\bf r}_2) \rangle = 0
\end{equation}
which implies that the dominant term for $|{\bf r}_1-{\bf r}_2|$ large
in this expression must vanish as well. Hence
\begin{equation}
   \alpha_S \alpha_P \langle p({\bf r}_1) q({\bf r}_2) \rangle = 0.
\end{equation}
This ensures that $p({\bf r})$ and $q({\bf r})$ are {\sl different}
scaling operators sharing the same critical dimension $x$. Therefore
this magnetic critical dimension $x$ must have multiplicity 2. Note
that the second argument does not apply for energy-like operators like
$S_{i,j}S_{i+1,j}$, such that the energy scaling field will be
non-degenerate.

We performed MC simulations using the standard Metropolis algorithm,
again on the system with geometry $L\times M$, with $L$ varying from 5
to 12 and $M/L$ varying from 1.5 to 10. We sampled four different
versions of the stress tensor $t_{xx}({\bf r})$:
\begin{equation}
\begin{aligned}
   (1)\quad & \langle S_{i,j}S_{i+1,j} + P_{i,j}P_{i+1,j} -
                      S_{i,j}S_{i,j+1} - P_{i,j}P_{i,j+1} \rangle , \\
   (2)\quad & \langle S_{i,j}S_{i+2,j} + P_{i,j}P_{i+2,j} -
                      S_{i,j}S_{i,j+2} - P_{i,j}P_{i,j+2} \rangle , \\
   (3)\quad & \langle S_{i,j}P_{i,j}S_{i+1,j}P_{i+1,j} -
                      S_{i,j}P_{i,j}S_{i,j+1}P_{i,j+1}    \rangle , \\
   (4)\quad & \langle S_{i,j}P_{i,j}S_{i+2,j}P_{i+2,j} -
                      S_{i,j}P_{i,j}S_{i,j+2}P_{i,j+2}    \rangle .
\end{aligned}
\label{stressAT}
\end{equation}
Stress tensors (1) and (2) are defined such that the symmetry between
$S$ and $P$ spins is incorporated.

\begin{table}
   \caption{Monte Carlo results for the Ashkin-Teller model,
   corresponding to the six vertex model with Boltzmann weight
   $W=0.8$. The stress tensors (1) to (4) are defined in
   Eq.~(\ref{stressAT}). For notation see table~\ref{table_ising}.}
\label{table_AT}
\begin{tabular}{cd@{~}ld@{~}ld@{~}ld@{~}ll}
     &  \multicolumn{2}{c}{Stress tensor (1)} &
        \multicolumn{2}{c}{Stress tensor (2)} &
        \multicolumn{2}{c}{Stress tensor (3)} &
        \multicolumn{2}{c}{Stress tensor (4)} & Exact \\
\tableline
$\alpha$& 0.36  &(2)& 0.78  &(3)& 0.23  &(1)& 0.65  &(2)& \\
$c$     & 0.97  &(6)& 0.99  &(3)& 0.96  &(4)& 0.96  &(3)& 1 \\
$x_1$   & 0.128 &(3)& 0.130 &(2)& 0.128 &(3)& 0.128 &(2)& 0.125 \\
$x_2$   & 0.33  &(8)& 0.34  &(5)& 0.34  &(6)& 0.35  &(4)& 0.2908...\\
$x_3$   & 0.9   &(4)& 0.8   &(1)& 0.9   &(3)& 0.9   &(2)& 0.8596...\\
$\omega$& 3.8   &(2)& 4.1   &(1)& 4.0   &(2)& 4.2   &(1)& \\
g.o.f. & 0.80  &   & 0.016 &   & 0.76  &   & 0.065 &   & \\
\end{tabular}
\end{table}

It turns out that in this case three nontrivial dimensions have to be
included in the fit. This brings the total number of fit parameters to
no less than 10. Still, relatively good results are obtained; they are
summarized in table~\ref{table_AT}.

\subsection{The F-model}

A nice illustration of the importance of boundary conditions is
obtained when a dual version of the Ashkin-Teller model is considered.
As stated, the critical line of the Ashkin-Teller model can be mapped
exactly onto the F-model, using a duality transformation and a
graphical representation\cite{Baxter82}. On a finite system, however,
this mapping affects the boundary conditions, such that both models
with periodic boundary conditions will have a different behavior on the
torus.

The model we chose to consider in fact is an intermediate model between
the F-model and the Ashkin-Teller model, and is obtained from the
latter by applying a duality transformation on one of the spins $S$ or
$P$ only. In this way, we obtain two coupled Ising models, defined on
two interpenetrating sublattices. Both Ising models are equal; they
interact via a nearest-neighbor coupling such that a broken Ising bond
carries a Boltzmann weight $W$, where the weight $W$ is the same as the
$W$ in Eq.~(\ref{mapping}) of the Ashkin-Teller model. The coupling
between the Ising models only exists in the restriction that two broken
Ising bonds are not allowed to cross each other. An elementary square
of the lattice contains two spins of both sublattices; diagonally
opposed spins belong to the same sublattice. The restriction is that at
most one of the bonds over the elementary square may be broken.

The resulting model can easily be mapped on the F-model, seen as a
BCSOS height model\cite{Vanbeijeren77}. To this end, the Ising-Bloch
walls are identified with the steps, carried by the first four vertices
of the F-model. To become steps, walls have to be equipped with an
arrow; the steps have to be identified as a step up or a step down.
This arrow assignment is simply such that two adjacent Ising-Bloch
walls carry antiparallel arrows if they belong to the same sublattice,
and carry parallel arrows if they belong to different sublattices.

In this way, a configuration of the two Ising models is mapped onto a
configuration of the F-model, and vice versa. There is, however, a
difference in boundary conditions on the torus. If we consider the
F-model on a finite geometry as a height model, we have to allow for
defects at the boundary. The smallest defect in the F-model is a defect
of two unit heights, which corresponds to two steps running over the
system. The corresponding Ising configuration however, would have one
Ising-Bloch wall running over the system for each sublattice, which is
not allowed when the two Ising models have periodic boundary
conditions. Hence, for the F-model the allowed defects at the boundary
are height differences multiples of~2, whereas in the formulation of
the Ising models, the height differences at the boundary are multiples
of~4.

Related to these defects is a complication that arises, when one
naively tries to simulate this version of the F-model using a
single-spin Metropolis algorithm. As the updates in such an algorithm
are always local, it cannot generate configurations with defects around
the torus. The algorithm is able to generate islands of flipped spins,
but such an island never can cross an Ising-Bloch wall of the other
sublattice. This implies that the algorithm is non-ergodic; the part
of phase space it reaches is restricted to that part that has the same
defects at the boundary as the initial configuration.

That does not mean that the results of the simulation make no sense.
The model that results when using only the Metropolis algorithm is a
true height model, such that on the boundaries no defects are allowed
at all. This model renormalizes to the Gaussian model. The universal
form of its partition function is known\cite{Itzykson86}, but behaves
somewhat anomalously because it has a continuous spectrum of critical
dimensions, that result in an integral instead of a sum in 
Eq.~(\ref{fitexpression}). The universal partition sum of the Gaussian
model is the result of this integral. Inclusion of its form in our fit
for this model indeed yields the correct result.

The difficulty in boundary conditions, however, can easily be overcome
using a cluster algorithm, that allows for non-local updates of the
configurations. In our simulations, we used a standard Metropolis
algorithm for thermal equilibration, combined with a cluster
algorithm\cite{Swendsen87,Wolff89} that is able to generate defects, in
order to make sure that the whole phase space can be reached. We
performed simulations on the model with $L$ varying from 6 to 18 and
$M/L$ from 2 to 5. It turned out in this case the stress tensor reaches
its asymptotic value already for $M/L\approx5$.

We sampled two possible versions of the stress tensor,
\begin{equation}
   \begin{aligned}
   (1) \quad& \langle S_{i,j}S_{i+2,j  } - S_{i,j}S_{i  ,j+2}\rangle, \\
   (2) \quad& \langle S_{i,j}S_{i+3,j+1} - S_{i,j}S_{i-1,j+3}\rangle,
   \end{aligned}
   \label{stress-f}
\end{equation}
where we took into account that energy-like spin products always
must couple spins of the same sublattice. The most simple version of
the stress tensor couples nearest-neighbor spins of each sublattice,
but its expectation value on the system geometries that we considered
is zero by symmetry. Stress tensor (2) is, regarding its definition, a
mix of $t_{xx}({\bf r})$ and $t_{xy}({\bf r})$, but this is no problem
since, on the used geometry, any $t_{xy}({\bf r})$ is zero.

\begin{table}
   \caption{Monte Carlo results for the F-model with Boltzmann weight
   $W=0.8$. Two different stress tensors are used for the calculation
   of the central charge and critical dimensions. They are defined in
   Eq.~(\ref{stress-f}). For notation see table~\ref{table_ising}.}

\begin{tabular}{cd@{~}ld@{~}ll}
     &  \multicolumn{2}{c}{Stress tensor (1)} &
        \multicolumn{2}{c}{Stress tensor (2)} & Exact \\
\tableline
$\alpha$& 0.83  &(6)& 1.32  &(7)& \\
$c$     & 1.06  &(7)& 1.03  &(6)& 1 \\
$x_1$   & 0.291 &(6)& 0.289 &(5)& 0.2908... \\
$x_2$   & 0.7   &(1)& 0.65  &(9)& 0.8596...\\
$\omega$& 3.2   &(2)& 2.8   &(1)& \\
g.o.f. & 0.28  &   & 0.27  &   & \\
\end{tabular}

\label{f-model}
\end{table}
The fact that this model is a height model ensures that there are
basically two types of operators, spin wave and vortex operators, with
dimensions given in Eq.~(\ref{cgp}), that both are doubly degenerate.
(cf. \cite{Francesco87} for further discussion.) Hence, the lowest
critical dimensions have multiplicity 2. We fitted the resulting
expectation values of the different stress tensors, using two
non-trivial critical dimensions. The results are summarized in
table~\ref{f-model}.

It is noteworthy that the prefactor $\alpha$ in the definition of the
stress tensor is independent of the boundary conditions. Our fit on the
simulation that only used the Metropolis algorithm (described above)
yielded the same prefactors as those in table~\ref{f-model}. That means
that the expansion~(\ref{stressexpansion}) of the discrete stress
tensor in terms of scaling fields only depends on local properties.

It turned out that including values of $M/L$ smaller than 2 destroyed
the quality of the fit, yielding a far too low value of the `goodness
of fit'. The reason probably is that there are much more dimensions
$x_j$ present that are quite small and that start to become important
for values of $M/L$ smaller than 2. This can be seen from the value of
$x_2$ that follows from the fit; it is significantly lower than the
exact value of the second dimension. Apparently, in the fit program
$x_2$ plays the role of an `effective' dimension, incorporating the
values of several dimensions in one. This casts doubt on the validity
of the highest dimension that is given by the fit program, but is seen
not to affect the values of the central charge $c$ and the most
relevant dimension $x_1$.

\section{Simulation times and autocorrelations}

MC calculations of a marginal operator like the stress tensor typically
encounter additional difficulties as compared to observables like
energy and magnetization. The latter quantities have a relative error
in MC simulations that does not scale with the system size, whereas
this is not the case for an operator like the stress tensor; its
relative error increases with the system size.

This can be seen as follows: consider an operator $O({\bf r})$ of which
we want to calculate its expectation value. Its scaling behavior will
be dictated by a critical dimension $x$,
\begin{equation}
   \frac1{L^2}\sum_{\bf r} \langle O({\bf r}) \rangle \sim
      L^{-x},
\end{equation}
where $L$ is the linear system size. The error $\Delta_{O({\bf r})}$
in the average value is related to the number of samples $N$ in the MC
simulation and to the second moment of its distribution,
\begin{equation}
   \Delta_{O({\bf r})}^2 = \frac1{N} \frac1{L^4} \sum_{\bf r,r'}
      \langle O({\bf r}) O({\bf r}') \rangle -
      \langle O({\bf r}) \rangle\langle O({\bf r}') \rangle .
   \label{error}
\end{equation}
Note that $N$ stands for the number of statistically {\sl independent}
MC samples. The dependence on $L$ of the simulation time to reach
independent samples will be discussed below.

Typically, the double summation in the last expression has two
contributions; a short range and a long range contribution. The short
range contribution, say within a region with radius $R$, follows
\begin{equation}
   \sum_{|{\bf r}|<R}
	  \langle O({\bf r})                O(0) \rangle -
	  \langle O({\bf r}) \rangle\langle O(0) \rangle
   \rightarrow \text{constant},
\end{equation}
for $L$ large. The constant is roughly proportional to the radius $R$
when it is not too large. The long range contribution, on the other
hand, is dominated by the critical dimension $x$ as
\begin{equation}
   \sum_{|{\bf r}|>R}
	  \langle O({\bf r})                O(0) \rangle -
	  \langle O({\bf r}) \rangle\langle O(0) \rangle
   \sim
	  L^{2-2x},
\end{equation}
Now there are two cases. If the dimension $x\leq1$ the long range
contribution dominates Eq.~(\ref{error}), and the relative error in
$\langle O({\bf r})\rangle$ scales according to
\begin{equation}
   \frac{ \Delta_{O({\bf r})} }{ \langle O({\bf r}) \rangle }
   \sim \frac1{\sqrt{N}}.
\end{equation}
It is inversely proportional to the square root of the number of MC
samples, but does not scale with the system size $L$. This is the usual
case for, .e.g., magnetization and energy in the Ising model. In case
$x>1$, however, the short range contribution dominates the error for
large $L$, which implies that the relative error in $\langle O({\bf
r})\rangle$ scales according to
\begin{equation}
   \frac{ \Delta_{O({\bf r})} }{ \langle O({\bf r}) \rangle }
   \sim \frac1{\sqrt{N}} \; L^{x-1}.
   \label{samplesstress}
\end{equation}
We will want to obtain the same relative error for all different linear
system dimensions $L$ in our MC simulations. For observables having
$x\leq1$ this requires the same number of MC samples for all $L$. For
$x>1$, however, Eq.~(\ref{samplesstress}) dictates that $N\sim
L^{2x-2}$. In case of the stress tensor, having $x=2$, the number of MC
samples should thus be proportional to $L^2$.

At first sight, it seems that this fact makes it difficult to reach
large system sizes, as the simulation time is directly proportional to
the number of required MC samples. This, however, is only partly true.
The other parameter which determines the simulation time is the time it
takes to generate statistically independent configurations. Critical
systems are known to suffer from critical slowing down. If one uses the
standard Metropolis algorithm, the typical time $\tau$ it takes to
generate statistically independent configurations increases with the
system size as a power law.

Unexpectedly, it turns out that the stress tensor is remarkably
insensitive to critical slowing down. This can be judged from its
autocorrelation function. Let us define a MC cycle as one
attempted update per spin. The autocorrelation function of a certain
observable $O$ is defined as
\begin{equation}
   g(t) =
   \frac{
      \langle O_{t_0}O_{t_0+t} \rangle -
         \langle O_{t_0} \rangle^2
   }{
      \langle O_{t_0}^2 \rangle -
         \langle O_{t_0} \rangle^2
   }.
   \label{autocordef}
\end{equation}
The operator $O$ is, as usual, defined as $\sum_{\bf r}O({\bf r})$. Here 
$O_t$ denotes the value of $O$ after $t$ time steps, where a time
step is one cycle, i.e., one attempted update per spin. The
autocorrelation function $g(t)$ is normalized such that $g(0)=1$. In
practical situations, the number of MC cycles $t$ between two
consecutive MC samples has to be such that $g(t)$ is (almost) zero.

\begin{figure}
   \begin{center}
   \epsfig{file=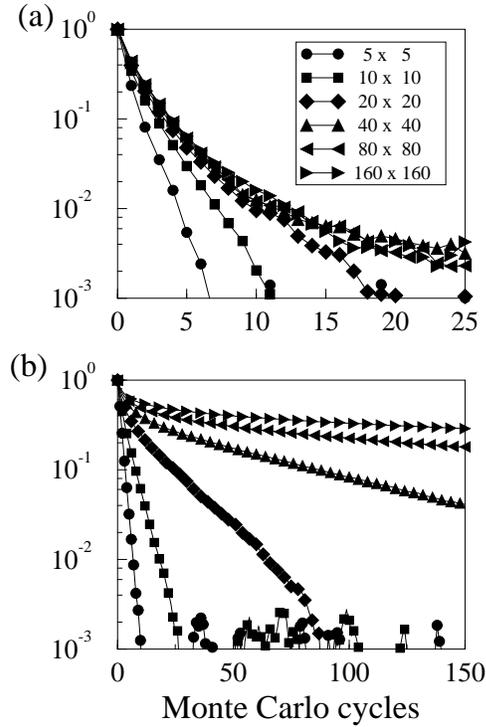,width=13cm}
   \end{center}
  \caption{Plots of the autocorrelation function $g(t)$ of
  Eq.~(\ref{autocordef}), where $t$ is the number of Monte Carlo
  cycles. Calculations are performed using the Metropolis algorithm
  for the Ising model at its critical point. System dimensions are
  indicated in the figure. (a) Autocorrelation function of the stress
  tensor. (b) Autocorrelation function of the energy. Note the
  difference in scale of the $x$-axes of the plots.}
  \label{autocorrelation}
\end{figure}

The observation that the stress tensor does not suffer very much from
critical slowing down follows from Fig.~\ref{autocorrelation}. Here we
plotted the autocorrelation functions $g(t)$ for the energy and the
stress tensor, in case of the Ising model at its critical point, for
several different system dimensions. For the number of cycles $t$ not
too small, the autocorrelation function of the energy shows a straight
line in the log-normal plot, meaning that its behavior is exponential
in $t$. Indeed, the behavior of the autocorrelation functions for
nearly critical systems is given by
\begin{equation}
   g(t) \sim \exp(-t/\tau) \quad\text{for $t$ large},
   \label{exp-behavior}
\end{equation}
where $\tau$ is the autocorrelation time. The dynamic scaling
hypothesis states that the time scale $\tau$ of a dynamical system is
connected with the length scale, which is the correlation length $\xi$,
and that this connection is described by a universal dynamic exponent
$z$,
\begin{equation}
   \tau \sim \xi^z.
\end{equation}
The exponent $z$ is believed to be connected to the dynamics of the
system (in our case, by the Metropolis algorithm) and to be the same
for all observables. For finite systems at their critical point, the
correlation length $\xi$ is bounded by the system dimension $L$, such
that
\begin{equation}
   \tau(L) \sim L^z.
   \label{finitesize}
\end{equation}
We extracted the values of $\tau(L)$, following from the
autocorrelation function of the energy in Fig.~\ref{autocorrelation},
by fitting the autocorrelation functions to Eq.~(\ref{exp-behavior}).
For this, we removed the first data points, up to the point where the
plot begins to show a straight line. The values of $\tau(L)$ were
fitted to Eq.~(\ref{finitesize}), yielding a value for $z$ of roughly
2. The quoted value in the literature\cite{Nightingale96} is
$z\approx2.17$, which is consistent with our findings.

The autocorrelation behavior of the stress tensor, however, is
dramatically different from that of the energy. Note the difference in
scale of the $t$-axes in Fig.~\ref{autocorrelation}. The
autocorrelation function of the stress tensor drops so sharply that the
exponential behavior can hardly, if at all, be seen. There is almost
no sign of critical slowing down; the autocorrelation functions even
seem to converge for larger and larger systems. Even for systems as
large as $180\times180$ spins the autocorrelation function behaves not
significantly different from smaller system sizes.

\begin{figure}
   \begin{center}
   \epsfig{file=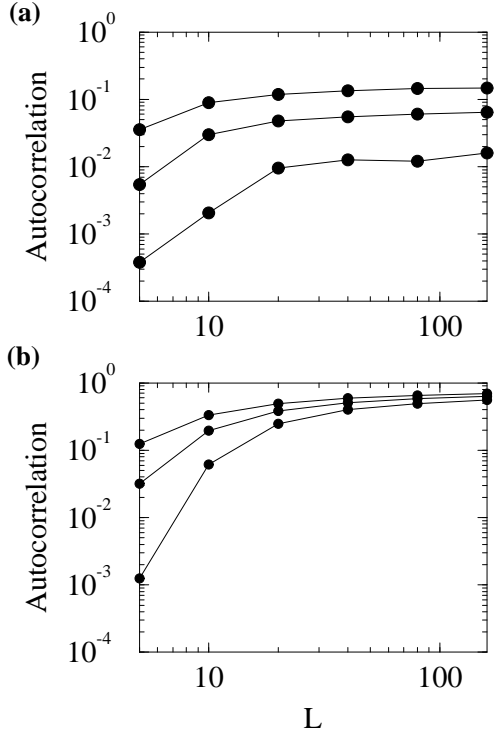,width=13cm}
   \end{center}
   \caption{Plots of the autocorrelation function $g(t)$ of
   Eq.~(\ref{autocordef}) versus the system size $L$. From low to high,
   the plots amount to $t=3$, $t=5$ and $t=10$. (a) $g(t)$ for
   the stress tensor. (b) $g(t)$ for the energy. The plots show that for
   the energy, $g(t)$ converges to a value independent of $t$, which
   must be~1. For the stress tensor, however, $g(t)$ converges to a
   value that does depend on $t$.}
   \label{converge}
\end{figure}

These findings can be explained as follows. The dynamic scaling
hypothesis in its general form considers the combined spatial and
timecorrelation function $G({\bf r},t)$, defined as
\begin{equation}
   G({\bf r},t) = 
	 \langle O({\bf r}_0,t_0) O({\bf r}_0+{\bf r},t_0+t) \rangle -
	 \langle O({\bf r}_0,t_0) \rangle^2,
\end{equation}
for a certain operator $O({\bf r},t)$. Here the dynamics of the system
is explicitly taken into account by the time dependence of the operator.
The dynamic scaling hypothesis states that
\begin{equation}
   G({\bf r},t) = b^{-2x} G(b^{-1} {\bf r},b^{-z} t),
   \label{dsh}
\end{equation}
where $x$ is the critical dimension of the operator $O({\bf r},t)$. In
terms of this correlation function, the autocorrelation function $g(t)$
of Eq.~(\ref{autocordef}) can be expressed as
\begin{equation}
   g(t) = 
	  \frac{
		 \int_0^L d^2{\bf r} \; G({\bf r},t)
	  }{
		 \int_0^L d^2{\bf r} \; G({\bf r},0)
	  }.
   \label{gt}
\end{equation}
The integral is over the finite volume $L^2$. The dynamic scaling
hypothesis~(\ref{dsh}) will be valid provided that the appearing lengths
are smaller than the correlation length $\xi$, and the times are smaller
than the autocorrelation time $\tau$, given by $\tau\sim\xi^z$. For
finite systems, $\xi\sim L$. In that case, Eq.~(\ref{dsh}) can be
rephrased to
\begin{equation}
   G({\bf r},t) = t^{-2x/z} G(t^{-1/z}{\bf r},1),
\end{equation}
which yields the $L$- and $t$-dependence in the scaling limit of
Eq.~(\ref{gt}). Using~(\ref{dsh}), the nominator is
\begin{equation}
   t^{(2-2x)/z} \int_0^{t^{-1/z}L} d^2{\bf r} \; G({\bf r},1),
\end{equation}
and $G({\bf r},1)$ must follow the usual spatial behavior 
$|{\bf r}|^{-2x}$. Now the scaling behavior of the integral depends on
whether it converges or diverges for large $L$. Making this distinction,
the scaling behavior of Eq.~(\ref{gt}) is
\begin{equation}
\begin{aligned}
   g(t) & \sim \text{constant} & \quad\text{for $x<1$}, \\
   g(t) & \sim t^{(2-2x)/z}    & \quad\text{for $x\ge1$}.
\end{aligned}
\end{equation}
This explains our MC results: both cases indicate that $g(t)$ must
become independent of $L$ in the scaling limit, i.e., for large $L$.
The case of the energy, having $x=1$, states that $g(t)$ must converge
to a value independent of $t$, whereas the case of the stress tensor
implies that $g(t)$ becomes a true power law in $t$. This behavior
indeed can be seen in Fig.~\ref{converge}, where for the stress tensor
and for the energy, the values of $g(t)$ are plotted as a function of
system size $L$ for several values of $t$. The plot for the energy
indicates that $g(t)$ converges to 1, whereas the asymptote of $g(t)$
for the stress tensor is seen to depend on $t$.

The above analysis also enables us to determine the scaling of the
typical simulation time of a MC simulation with the system size $L$.
This scaling will depend on the MC algorithm (i.e., on $z$) and on the
observable we want to know. Starting point is that we will want to
obtain the same relative error in the average value $\langle O({\bf
r})\rangle$ for each system dimension $L$. The error $\Delta_{O({\bf
r})}$ in the average is proportional to the second moment of the
correlation function,
\begin{equation}
   \Delta_{O({\bf r})}^2 = \frac1N \frac1{L^2} 
	  \int_0^{L^z}dt \int_0^{L} d^2{\bf r} \; G({\bf r},t).
\end{equation}
Here $N$ is the total number of MC cycles, which is supposed to be
much larger than the autocorrelation time $L^z$. Using, as above,
Eq.~(\ref{dsh}) and the distinction between converging and diverging
integrals with $L$, we obtain
\begin{equation}
\begin{aligned}
   \Delta_{O({\bf r})}^2 &\sim \frac1N L^{z-2x} 
	  \quad&\text{for $x<1+\frac12z$}&, \\
   \Delta_{O({\bf r})}^2 &\sim \frac1N L^{-2}
	  \quad&\text{for $x\ge1+\frac12z$}&.
\end{aligned}
\end{equation}
The relative error is obtained by dividing these values by $\langle
O({\bf r})\rangle$, which scales as $L^{-x}$. The typical number of MC
cycles $N$ is obtained by demanding it to be such that the same
relative error is obtained for all $L$. This yields
\begin{equation}
\begin{aligned}
  N &\sim L^z      &\quad\text{for $z>2x-2$}&, \\
  N &\sim L^{2x-2} &\quad\text{for $z\le2x-2$}&.
\end{aligned}
\end{equation}
This implies that for a relevant operator, like the energy in the Ising
model, faster convergence is obtained by a MC algorithm that has a
lower value of the dynamic exponent $z$. However, the case of the
stress tensor, $x=2$, represents a border case, because for the
Metropolis algorithm $z$ is only slightly larger than~2. This explains
why it is not necessary to use a more sophisticated cluster algorithm
for MC simulations on the stress tensor.

Note that the actual simulation {\sl time} is, in the case of a
Metropolis algorithm, proportional to $L^2N$, because the time needed
for a single MC cycle is simply proportional to the number of spins.
An important consequence of the above is that the typical simulation
time for the stress tensor is roughly proportional to $L^4$. This
contrasts with the computer time needed for transfer matrix
calculations, which is exponential in $L$.

Hence, in principal much larger system sizes can be reached with our MC
method than in the transfer matrix method, to calculate the central
charge. This is a promising conclusion for systems of which the value
of the central charge up to now is an open question\cite{Knops94}.

\section{Discussion and conclusion}

In this paper, we proposed a novel Monte Carlo technique for the
calculation of the central charge and some critical dimensions of
two-dimensional critical models. The technique is based on the
universal behavior of the stress tensor, an operator that plays an
important role in the theory of conformal invariance, but of which a
lattice representation can easily be found as well. The rough data,
following from the Monte Carlo simulation, require a decent fit program
to extract the central charge and critical dimensions. By comparing our
Monte Carlo analysis for three different models with their exact
results, we show that the method works. We explain why, on one hand,
simulations on the stress tensor are difficult because its expectation
value is equipped with larger error bars than usual. On the other hand,
it turns out that the simulations are much easier than usual because
the stress tensor shows to be remarkably insensitive to critical
slowing down. The latter observation notably ensures that the typical
simulation time of our method scales with the system size $L$ roughly
as $L^4$, in contrast with transfer matrix calculations, which scale
exponentially as $n^L$, where $n$ is the number of different spin
states.

Hence, in principal much larger system sizes can be reached with the
proposed method than with transfer matrix calculations. For that
reason, we expect the merits of our method to lie mainly in simulations
on models with a large number of spin states $n$, especially when these
states become continuous, as, e.g., in the $XY$-Ising model. As the
stress tensor is highly insensitive to critical slowing down,
advantages can also be obtained when no cluster algorithm is available
for the Monte Carlo simulations.

\begin{acknowledgments}
It is a pleasure to thank Henk Bl\"ote for discussions on
dynamical systems.
\end{acknowledgments}

\end{document}